\documentclass[aps,pr,twocolumn,
superscriptaddress,groupedaddress,nofootinbib,floatfix]{revtex4}  

\usepackage{graphicx}  
\usepackage{dcolumn}   
\usepackage{bm}        
\usepackage{amssymb}  
\usepackage{amssymb}
\usepackage{amsxtra}
\usepackage{euscript} 
\usepackage{amsthm}
\usepackage{color}
\usepackage{ulem}

\begin{document}

\title{Hints of (de)confinement in Yang-Mills-Chern-Simons theories in the maximal Abelian gauge}

\author{Luigi C. Ferreira}
\email{luigicferreira@gmail.com }
\affiliation{Instituto de F\'isica, Universidade Federal Fluminese, Av. Litor\^anea s/n, 24210-346, Niter\'oi, RJ, Brazil}

\author{Antonio D. Pereira}
\email{adpjunior@id.uff.br}
\affiliation{Instituto de F\'isica, Universidade Federal Fluminese, Av. Litor\^anea s/n, 24210-346, Niter\'oi, RJ, Brazil}

\author{Rodrigo F. Sobreiro}
\email{rodrigo_sobreiro@id.uff.br}
\affiliation{Instituto de F\'isica, Universidade Federal Fluminese, Av. Litor\^anea s/n, 24210-346, Niter\'oi, RJ, Brazil}

\begin{abstract}
\noindent The study of Yang-Mills theories in three dimensions is an insightful playground to grasp important features for the four-dimensional case. Additionally, in three dimensions, the Chern-Simons term can be introduced with a mass parameter of topological nature. Quantizing such a theory in the continuum demands a gauge fixing which, in general, is plagued by Gribov copies. In this work, Yang-Mills-Chern-Simons theories are quantized in the maximal Abelian gauge and the existence of infinitesimal Gribov copies is taken into account. The elimination of copies modifies the (Abelian) gluon propagator leading to two different phases: one in which all poles are complex and thus interpreted as a confining phase and another where an excitation which can be part of the physical spectrum is present. 
\end{abstract}

\maketitle

\section{Introduction}

Yang-Mills theories are the main building blocks of the successful Standard Model of particle physics. The description of fundamental interactions is very well accommodated in the framework of non-Abelian gauge theories. Despite of the great understanding of these theories in the perturbative regime, many challenging open problems are present when they become strongly correlated. Due to the inefficiency of perturbation theory at this regime, non-perturbative tools are required and a systematic description of the theory becomes much more challenging. Several different non-perturbative frameworks are available and the most fruitful avenue seems to seek for an interplay between them. 

One of the most iconic non-perturbative phenomena which begs for an analytic consistent description is color confinement. Although several complementary hints try to complete the patchwork for a full satisfactory description of confinement, a comprehensive mechanism still lacks. Different approaches as functional methods, effective models, lattice simulations and holographic techniques  \cite{Greensite:2011zz,Brambilla:2014jmp,Alkofer:2000wg,Binosi:2009qm,Tissier:2010ts,Maas:2011se,Siringo:2015wtx,Cyrol:2016tym,Huber:2018ned,Oxman:2018dzp,Reinhardt:2017pyr,Witten:1998zw} are used as complementary perspectives on the problem. Another approach which can provide new insights towards a mechanism that describes confinement comes from the fact that, at non-perturbative regimes, the standard quantization procedure for non-Abelian gauge theories requires a modification. The reason behind that are the so-called Gribov copies. The quantization of gauge theories in continuum space(time) requires a gauge-fixing procedure, i.e., a selection of a gauge field per gauge orbit. In principle, this is achieved by imposing a constraint to the gauge field. However, in the seminal works \cite{Gribov:1977wm,Singer:1978dk}, it was shown that an ideal gauge-fixing condition, i.e., the selection of only one representative per gauge orbit, does not exist in non-Abelian gauge theories for continuous and covariant gauges. The existence of field configurations which satisfy the same gauge condition {\it and} are related by a gauge transformation is the so-called {\it Gribov problem}, see \cite{Vandersickel:2012tz,Sobreiro:2005ec,Pereira:2016inn} for reviews. Such spurious configurations are dubbed Gribov copies.

One of the assumptions of the Faddeev-Popov gauge-fixing procedure is that the gauge-fixing condition is ideal, i.e., only one representative per gauge orbit satisfies the gauge condition. Thus, Gribov copies entail a breakdown of the Faddeev-Popov prescription. Nevertheless, at the perturbative regime, i.e., quantum fluctuations around the trivial vacuum $A^a_\mu = 0$, where $A^a_\mu$ is the gauge field, such redundant configurations do not play any role. In fact, one can show that the Faddeev-Popov procedure is well-grounded at the perturbative level. However, this is not true anymore at the non-perturbative regime where fluctuations around the trivial vacuum are not small. Hence, it is conceivable that Gribov copies must be appropriately taken into account by modifying the Faddeev-Popov prescription non-perturbatively. A consistent quantization of Yang-Mills theories which is valid at the strongly coupled regime seems to require to deal with Gribov copies. As already pointed out in \cite{Gribov:1977wm}, a modification in the quantization of non-Abelian gauge theories might be key for the understanding of the mechanism behind confinement. 

In \cite{Gribov:1977wm}, it was proposed to restrict the path integral measure gauge-fixed to the Landau gauge to a region, the {\it Gribov region}, which is free of infinitesimal Gribov copies, i.e., those generated by infinitesimal gauge transformations. This region features several important properties as, e.g., all gauge orbits cross it once and thereby the restriction does not exclude any physical configuration from the configuration space \cite{DellAntonio:1991mms}. The restriction can be effectively implemented by the introduction of a non-local term, the {\it horizon function}, to the classical action. This was worked out at leading order in \cite{Gribov:1977wm} and generalized to all orders in \cite{Zwanziger:1989mf} using a slightly different method. The equivalence of the prescriptions was proved in \cite{Capri:2012wx}. The resulting action can be localized by the introduction of a suitable set of auxiliary fields giving rise to the so-called {\it Gribov-Zwanziger action}. It is local, perturbatively renormalizable at all orders and effectively implements the restriction of the functional measure to the Gribov region, thus excluding the infinitesimal Gribov copies. Large copies are still present within the Gribov region \cite{vanBaal:1991zw} and a complete elimination of them requires a further restriction to the {\it fundamental modular region}. A consistent restriction of the functional measure to this region is not known so far, see for instance \cite{vanBaal:1991zw} and references therein.

The tree-level gluon propagator derived from the Gribov-Zwanziger action is suppressed in the infrared (IR) and attains vanishing value at zero momentum. This implies a violation of reflection positivity, a fact often interpreted as a signal of confinement since they cannot be associated with physical excitations in the spectrum of the theory. The ghost propagator is enhanced in the IR and diverges as $\propto1/k^4$ for low momenta. More recent gauge-fixed lattice simulations, however, point to a finite gluon propagator at vanishing momentum and a non-enhanced ghost propagator in the IR in the Landau gauge \cite{Cucchieri:2007rg,Maas:2008ri,Bogolubsky:2009dc}. In \cite{Dudal:2007cw,Dudal:2008sp,Dudal:2011gd} it was realized that the restriction to the Gribov region leads to further non-perturbative effects that must be taken into account. In particular, the auxiliary fields introduced to localize the horizon function acquire their own dynamics and give rise to the formation of dimension-two condensates. Thence, a renormalizable and local action which takes into account the restriction to the Gribov region and further non-perturbative effects as the formation of condensates was proposed, the {\it refined Gribov-Zwanziger} action. The resulting gluon propagator attains a finite value at zero momentum and the ghost propagator is not enhanced in the IR, in agreement with the most recent lattice data. 

Despite of being intrinsically associated with the non-trivial bundle structure of non-Abelian gauge theories, the (partial) solution of the Gribov problem is intimately related with the specific choice of gauge-fixing. As an illustration, the elimination of infinitesimal Gribov copies in the Landau gauge relies of the fact that the Faddeev-Popov operator is Hermitian, a property which does not hold in general gauges. Another popular choice for gauge condition is the so-called maximal Abelian gauge (MAG) \cite{tHooft:1981bkw,Kronfeld:1987vd,Kronfeld:1987ri}. In this gauge, non-Abelian and Abelian components of the gauge field satisfy different gauge conditions. For concreteness, we consider the gauge group as being $SU(2)$ and the MAG is defined by
\begin{equation}
\EuScript{D}^{ab}_{\mu}A^{b}_{\mu}=0\,,\qquad\mathrm{and}\qquad\partial_\mu A_{\mu}=0\,,
\label{intro1}
\end{equation}
where $A^{a}_{\mu}$ are the non-Abelian components of the gauge field and $A_\mu$ is the Abelian component. The lower-case Latin indices run from $a=1,2$. The covariant derivative $\EuScript{D}^{ab}_{\mu}\equiv \delta^{ab}\partial_\mu -g\epsilon^{ab}A_{\mu}$ is taken with respect to the Abelian field. The coupling constant is denoted by $g$ and $\epsilon^{ab}$ is the totally antisymmetric Levi-Civita in two dimensions. This gauge features a Hermitian Faddeev-Popov operator and the implementation of a solution akin to the Gribov-Zwanziger one in the Landau gauge is viable. The implementation of the restriction of the path integral to a Gribov region for the MAG was investigated in several works, see, e.g., \cite{Capri:2005tj,Dudal:2006ib,Capri:2006cz,Capri:2008ak,Capri:2008vk,Capri:2010an,Guimaraes:2011sf,Capri:2013vka,Gongyo:2013rua}. Remarkably, the effects of the Gribov copies modify the Abelian gluon propagator giving rise to the violation of reflection positivity. For the non-Abelian sector, the formation of dimension-two condensates leads to a Yukawa-like behavior. For a sufficiently large value for such a mass parameter, we have the realization of the Abelian dominance mechanism at low energies, \cite{Kronfeld:1987vd,Kronfeld:1987ri}. In general, however, the Faddeev-Popov operator is not Hermitian and the very definition of a Gribov region becomes unclear. Yet, progress was made in this direction in recent years for linear covariant gauges, Landau-MAG interpolating gauge and Curci-Ferrari gauges, see, e.g., \cite{Sobreiro:2005vn,Capri:2015pja,Capri:2015ixa,Capri:2015nzw,Pereira:2013aza,Pereira:2014apa,Serreau:2013ila,Serreau:2015yna,Pereira:2016fpn,Capri:2018ijg}. See also \cite{Lavrov:2013boa,Moshin:2014xka}.

A common strategy to gain insights for four-dimensional Yang-Mills theories is to analyze the theory is three dimensions. In this case, the theory is sufficiently non-trivial, i.e., it is confining, but it features simplifications as it is super-renormalizable. Moreover, its correlation functions in the Landau gauge qualitatively agree with those computed in four dimensions. See, e.g., \cite{Dudal:2008rm,Cucchieri:2016jwg,Huber:2016tvc,Corell:2018yil,Junior:2019fty} for some recent references on three-dimensional Yang-Mills theories. Moreover, in three dimensions, the gauge field can acquire a mass of topological nature due to the introduction of the Chern-Simons term, see \cite{Deser:1981wh,Deser:1982vy}. The Chern-Simons term is parity and time-reversal odd and therefore, it is not generated radiatively in pure Yang-Mills theories. It breaks large gauge transformations invariance unless a quantization rule is assigned to the mass parameter. However, it is invariant under infinitesimal gauge transformations. Hence, Yang-Mills-Chern-Simons theories where the Chern-Simons term is added to the Yang-Mills action provides a description of massive gauge fields and features, at least, infinitesimal gauge symmetry. In \cite{Canfora:2013zza}, the path-integral quantization of such system was revisited in light of the existence of Gribov copies in the Landau gauge. The Gribov parameter and the topological mass compete showing a transition from a phase where all poles of the gauge field propagator are not physical and therefore hints to a confining phase to a deconfining one where physical excitations are present in the spectrum. Qualitatively, this emulates the behavior of Yang-Mills theories in the presence of Higgs fields when Gribov copies are taken into account as discussed in \cite{Capri:2012cr,Capri:2012ah,Capri:2013gha,Capri:2013oja}. In \cite{Gomez:2015aba}, Higgs fields were added to Yang-Mills-Chern-Simons theories in the Landau gauge. 

In this work, we investigate the possibility of existence of different phases in Yang-Mills-Chern-Simons theories quantized in the MAG for $SU(2)$ gauge group. The main motivation lies on the fact the non-linear character of the MAG induces new interactions between matter fields and Faddeev-Popov ghosts making the analysis of Yang-Mills-Higgs systems in this gauge much more non-trivial, \cite{Capri:2015pxa}. In the case of Yang-Mills-Chern-Simons theories, we can explore qualitative features of such phase diagram without introducing new fields in the MAG. Moreover, this gauge provides the opportunity to explore the Abelian dominance mechanism more explicitly. Finally, as a technical motivation, the properties of the Gribov region in the MAG are well known and we can easily import the already developed technology for their elimination in four-dimensional Yang-Mills theories to this system.  

The paper is organized as follows: in Sect.~\ref{pre} we provide a short review of the Gribov problem in the Landau gauge. In Sect.~III we set up the conventions of Yang-Mills-Chern-Simons theories in the MAG and discuss the tree-level propagators without the elimination of Gribov copies. Sect.~IV is devoted to the elimination of Gribov copies in Yang-Mills-Chern-Simons theories in the MAG, which entails a modification to the gauge-field propagator. In Sect.~V we discuss the analytic properties of the new propagators and show that there are different phases (confining and deconfining) for different values of the underlying mass parameters of the theory. Finally, we collect some perspectives and conclusions. 

\section{The Gribov problem in a nutshell}\label{pre}

The perturbative quantization of Yang-Mills theories in a continuum setting requires a gauge-fixing procedure. In the path integral formulation, this is typically achieved by the so-called Faddeev-Popov procedure. A fundamental assumption in this method is the existence of a single representative of the gauge field per gauge orbit that satisfies the gauge condition, i.e., if $A^a_\mu$ and $A^{\prime\,a}_\mu$ are field configurations that satisfy the gauge condition $F^a[A]=0$, then they are not related by a gauge transformation, namely
\begin{equation}
A^{\prime}_\mu \neq UA_\mu U^\dagger + \frac{i}{g}U\partial_\mu U^\dagger\,,
\label{gp1}
\end{equation}
with $U$ and element of the gauge group\footnote{For concreteness, we take the gauge group to be $SU(N)$. Later on, we shall restrict to $SU(2)$.} and $g$ is the gauge coupling constant. However, as first discussed in \cite{Gribov:1977wm}, standard gauge-fixing conditions do not satisfy this requirement and in \cite{Singer:1978dk}, it was shown that this is a rather general issue than a simple pathology in particular gauge conditions. 

As a particular example, consider the Landau gauge where $\partial_\mu A^A_\mu=0$ where upper case Latin indices run through $1,2,\ldots,N^2-1$. If this condition is ideal, i.e., picks just one representative per orbit, taking a gauge field configuration $A^{\prime\,A}_\mu$ which is related to $A^{A}_\mu$ by a gauge transformation yields $\partial_\mu A^{\prime\,A}_\mu \neq 0$. This can be checked explicitly, e.g., taking an infinitesimal gauge transformation, i.e.,
\begin{equation}
A^{\prime\,A}_\mu = A^{A}_\mu-D^{AB}_\mu \xi^B\,,
\label{gp2}
\end{equation} 
with $D^{AB}_{\mu}=\delta^{AB}\partial_\mu - gf^{ABC}A^C_\mu$ denoting the covariant derivative in the adjoint representation of the gauge group, $f^{ABC}$ being the structure constants and $\xi^A$, an infinitesimal gauge parameter. Taking the divergence of \eqref{gp2} and imposing the gauge condition on $A^A_\mu$, leads to
\begin{equation}
\partial_\mu A^{\prime\,A}_\mu = -\partial_\mu D^{AB}_\mu \xi^B\,,
\label{gp3}
\end{equation}
and becomes clear that $\partial_\mu A^{\prime\,A}_\mu \neq 0$ only if the operator $\EuScript{M}^{AB}=-\partial_\mu D^{AB}_\mu$, the Faddeev-Popov operator, does not develop zero modes. It turns out that the Faddeev-Popov operator has zero modes and, thus, the Landau gauge condition picks up more than one representative per gauge orbit. Such spurious configurations are the so-called Gribov copies and their existence is the well-known Gribov problem. In fact, this argument is restricted to copies that are connected to a field configuration by infinitesimal gauge transformations and should be called ``infinitesimal copies''. It is possible to show that copies are also generated by finite gauge transformations  \cite{Gribov:1977wm,Henyey:1978qd,vanBaal:1991zw}. This means that the standard assumption in the implementation of the Faddeev-Popov trick of the non-existence of more than one representative per gauge orbit is not fulfilled.

To improve the gauge-fixing procedure, Gribov proposed to restrict the configuration space of gauge fields to a region $\Omega$, the Gribov region, defined by
\begin{equation}
\Omega = \left\{A^{A}_{\mu},\,\,\,\partial_\mu A^{A}_{\mu}=0\,|\,\EuScript{M}^{AB}>0\right\}\,.
\label{gp33}
\end{equation} 
Since, in the Landau gauge, $\EuScript{M}^{AB}$ is a Hermitian operator, then imposing the functional integration to a domain where it is positive becomes a meaningful task. Such a region features important properties: it is bounded in all directions in field space, it is convex and all gauge orbits cross it at least once, see \cite{DellAntonio:1989wae,DellAntonio:1991mms}. It is not free of Gribov copies. There are still those generated by finite gauge transformation. In order to eliminate all copies, one should restrict the functional integral domain of integration to the so-called fundamental modular region which is free of copies by definition and it is a subspace of the Gribov region. However, a practical implementation of the restriction to this region is still an open problem.

Formally, the path integral for Yang-Mills theories restricted to the Gribov region is expressed as
\begin{equation}
\EuScript{Z}_{\mathrm{YM}}=\int_{\Omega}\left[\EuScript{D}\mu_{\mathrm{YM}}\right]\,\mathrm{e}^{-S_{\mathrm{YM}}-S_{\mathrm{FP}}}\,,
\label{gp4}
\end{equation}
with
\begin{equation}
S_{\mathrm{YM}} = \frac{1}{4}\int_x~F^{A}_{\mu\nu}F^{A}_{\mu\nu}\,,
\label{gp5}
\end{equation}
and
\begin{equation}
S_{\mathrm{FP}} = \int_x\left(b^A\partial_\mu A^A_\mu + \bar{c}^{A}\partial_\mu D^{AB}_{\mu}c^B\right)\,,
\label{gp6}
\end{equation}
with $\int_x=\int \mathrm{d}^dx$ and $\left[\EuScript{D}\mu_{\mathrm{YM}}\right] = \left[\EuScript{D}A\right]\left[\EuScript{D}b\right]\left[\EuScript{D}\bar{c}\right]\left[\EuScript{D}c\right]$. The fields $b^A$, $\bar{c}^A$ and $c^A$ denote, respectively, the Nakanishi-Lautrup field and Faddeev-Popov ghosts. The field-strength $F^A_{\mu\nu}$ is given by $F^A_{\mu\nu} = \partial_\mu A^A_\nu-\partial_\nu A^A_\mu+gf^{ABC}A^B_\mu A^C_\nu$. The restriction to $\Omega$ can be effectively implemented by a modification of the measure which in turn can be lifted to the Boltzman factor. This was worked out at leading order in $g$ by Gribov \cite{Gribov:1977wm} and extended to all orders by Zwanziger in \cite{Zwanziger:1989mf} using different methods. The result of these two different approaches is the same \cite{Capri:2012wx}. Therefore, the functional integral restricted to the Gribov region can be expressed as
\begin{equation}
\EuScript{Z}_{\Omega}=\int\left[\EuScript{D}\mu_{\mathrm{YM}}\right] \,\mathrm{e}^{-S_{\mathrm{YM}}-S_{\mathrm{FP}}-\gamma^4 H(A)+\gamma^4 dV(N^2-1)}\,,
\label{gp7}
\end{equation}
where $V$ corresponds to the volume of spacetime and $\gamma$ is a mass parameter known as the Gribov parameter. The function $H(A)$ is the so-called horizon function and is written as
\begin{equation}
H(A)=g^2\int_{x,y}~f^{ABC}A^{B}_{\mu}(x)\left[\EuScript{M}^{-1}(A)\right]^{AD}_{x,y}f^{DEC}A^{E}_{\mu}(y)\,.
\label{gp8}
\end{equation}
The Gribov parameter $\gamma$ is not free but fixed by a gap equation,
\begin{equation}
\langle H(A)\rangle=dV(N^2-1)\,,
\label{gp9}
\end{equation}
where $\langle\ldots\rangle$ is computed with respect to the modified partition function \eqref{gp7}. The horizon function is non-local and thereby the resulting action which implements the restriction to the Gribov region is non-local. Such a non-locality can be cured by the introduction of auxiliary fields. This procedure leads to
\begin{equation}
\EuScript{Z}_{\mathrm{GZ}}=\int\left[\EuScript{D}\mu_{\mathrm{GZ}}\right] \,\mathrm{e}^{-S_{\mathrm{GZ}}}\,,
\label{gp10}
\end{equation}
with
\begin{eqnarray}
S_{\mathrm{GZ}}&=&S_{\mathrm{YM}}+S_{\mathrm{FP}}\nonumber\\
&-&\int_x\left(\bar{\varphi}^{AC}_{\mu}\EuScript{M}^{AB}\varphi^{BC}_{\mu}-\bar{\omega}^{AC}_{\mu}\EuScript{M}^{AB}\omega^{BC}_{\mu}\right)\nonumber\\
&+&\gamma^2\int_x~gf^{ABC}A^{A}_{\mu}(\bar{\varphi}+\varphi)^{BC}_{\mu}\nonumber\\
&-&\gamma^4\int_xd(N^2-1)\,,
\label{gp11}
\end{eqnarray}
and 
\begin{equation}
\left[\EuScript{D}\mu_{\mathrm{GZ}}\right] =\left[\EuScript{D}A\right]\left[\EuScript{D}b\right]\left[\EuScript{D}\bar{c}\right]\left[\EuScript{D}c\right]\left[\EuScript{D}\bar{\varphi}\right]\left[\EuScript{D}\varphi\right]\left[\EuScript{D}\bar{\omega}\right]\left[\EuScript{D}\omega\right]\,. 
\label{gp12}
\end{equation}
Action \eqref{gp11} is known as the Gribov-Zwanziger action. It implements the restriction of the path integral measure to $\Omega$ in a local and renormalizable way, see \cite{Zwanziger:1989mf}. 

From the previous discussion, it is clear that the elimination of (infinitesimal) Gribov copies relied on some particular properties of the gauge condition. In particular, the hermiticity of the Faddeev-Popov operator holds for the Landau gauge, but it is not a general property. There are other gauges which feature a Hermitian Faddeev-Popov operator. The MAG is an example and the analogous analysis \textit{a la} Gribov and Zwanziger can be repeated. This was done in \cite{Capri:2005tj,Dudal:2006ib,Capri:2006cz,Capri:2008ak,Capri:2008vk,Capri:2010an,Guimaraes:2011sf,Capri:2013vka,Gongyo:2013rua}. Conversely, for gauges where the Faddeev-Popov operator is not Hermitian, some progress has been achieved over the last years, see, e.g. \cite{Sobreiro:2005vn,Capri:2015pja,Capri:2015ixa,Capri:2015nzw,Pereira:2013aza,Pereira:2014apa,Serreau:2013ila,Serreau:2015yna,Pereira:2016fpn,Capri:2018ijg}. Another important observation is that the restriction to the Gribov region does not rely on the form of the gauge-invariant action we want to gauge fix and on the spacetime dimension $d$. This means that for three-dimensional Yang-Mills theories, the analysis is exactly the same. This will be explored in the next sections in the case of the MAG.

\section{Yang-Mills-Chern-Simons theories in the maximal Abelian gauge}

The starting point of the present analysis is the Yang-Mills-Chern-Simons action defined in three Euclidean dimensions ($d=3$) by the action,
\begin{equation}
S_{\mathrm{YMCS}} = S_{\mathrm{YM}} + S_{\mathrm{CS}}\,,
\label{ymcsmag1}
\end{equation}
with $S_{\mathrm{YM}}$ given by \eqref{gp5} with $d=3$ and
\begin{equation}
S_{\mathrm{CS}} = -iM\int_x~\epsilon_{\mu\rho\nu}\left(\frac{1}{2}A^{A}_{\mu}\partial_\rho A^{A}_{\nu}+\frac{g}{3!}f^{ABC}A^{A}_{\mu}A^{B}_{\rho}A^{C}_{\nu}\right)\,.
\label{ymcsmag2}
\end{equation}
The parameter $M$ is a mass parameter of topological origin and $\epsilon_{\mu\rho\nu}$ is the totally anti-symmetric Levi-Civita symbol. The action \eqref{ymcsmag1} is invariant under transformations \eqref{gp2}. For concreteness, we restrict from now on the gauge group to be $SU(2)$. In this work, we are interested to quantize the theory defined by \eqref{ymcsmag1} in the MAG. In order to introduce the gauge-fixing action and the correspondent Faddeev-Popov ghosts, we employ the Cartan decomposition, i.e., we decompose the gauge field in Abelian and non-Abelian components,
\begin{equation}
A^{A}_{\mu}T^{A} = A^{a}_{\mu}T^a + A^3_{\mu}T^3\equiv A^{a}_{\mu}T^a + A_{\mu}T\,,
\label{ymcsmag3}
\end{equation}
with $T^{A}$ being the generators of $SU(2)$ and $a=\left\{1,2\right\}$ correspond to the non-Abelian components. The generator $T^3\equiv T$ commutes with all the others and $A^{3}_{\mu}=A_\mu$ is the Abelian component of the gauge field. When decomposed, the structure constants $f^{ABC}$ are non-vanishing when two indices are non-Abelian and the third is the Abelian component. For simplicity, we employ the  normalization $f^{ab3}=\epsilon^{ab}$. After decomposition \eqref{ymcsmag3}, the action \eqref{ymcsmag1} is expressed as
\begin{eqnarray}
S_{\mathrm{YMCS}}&=&\frac{1}{4}\int_x \left(F^{a}_{\mu\nu}F^{a}_{\mu\nu}+F_{\mu\nu}F_{\mu\nu}\right)\nonumber\\
&-&iM\int_x~\epsilon_{\mu\rho\nu}\left(\frac{1}{2}A^{a}_\mu\partial_\rho A^{a}_{\nu}+\frac{1}{2}A_\mu\partial_\rho A_{\nu}\right)\nonumber\\
&-&iM\int_x~\epsilon_{\mu\rho\nu}\frac{g}{2}\epsilon^{ab}A^{a}_\mu A^{b}_{\rho}A_\nu\,,
\label{ymcsmag4}
\end{eqnarray}
with
\begin{eqnarray}
F^{a}_{\mu\nu} &=& \EuScript{D}^{ab}_\mu A^{b}_{\nu}-\EuScript{D}^{ab}_\nu A^{b}_{\mu}\nonumber\\
F_{\mu\nu} &=& \partial_\mu A_\nu - \partial_\nu A_\mu +g\epsilon^{ab}A^{a}_{\mu}A^{b}_{\nu}\,.
\label{ymcsmag5}
\end{eqnarray}
The covariant derivative $\EuScript{D}^{ab}_{\mu}\equiv \delta^{ab}\partial_\mu -gf\epsilon^{ab}A_{\mu}$ is defined in terms of the Abelian component of the gauge field.

In order to implement the MAG condition in \eqref{ymcsmag4}, we have to introduce a BRST-exact\footnote{See Ap.~\ref{appendix} for more details on the BRST transformations. } term given by
\begin{eqnarray}
S^{\mathrm{MAG}}_{\mathrm{FP}}&=&s\int_x~\left(\bar{c}^a \EuScript{D}^{ab}_{\mu}A^{b}_{\mu}+\bar{c}\,\partial_\mu A_\mu\right)\nonumber\\
&=&\int_x\left[ib^a\EuScript{D}^{ab}_{\mu}A^{b}_{\mu}-\bar{c}^{a}\mathcal{M}^{ab}c^b+g\epsilon^{ab}\bar{c}^a\left(\EuScript{D}^{bc}_{\mu}A^{c}_{\mu}\right)c\right.\nonumber\\
&+&\left.ib\,\partial_\mu A_\mu + \bar{c}\partial_\mu\left(\partial_\mu c + g\epsilon^{ab}A^{a}_{\mu}c^b\right)\right]\,,
\label{ymcsmag6}
\end{eqnarray}
where $\mathcal{M}^{ab}\equiv -\EuScript{D}^{ac}_{\mu}\EuScript{D}^{cb}_{\mu}-g^2\epsilon^{ac}\epsilon^{bd}A^{c}_{\mu}A^{d}_{\mu}$ is the Faddeev-Popov operator in the MAG. Hence, the quantization of the Yang-Mills-Chern-Simons in the MAG is given by the path integral 
\begin{equation}
\EuScript{Z}_{\mathrm{MAG}}=\int\left[\EuScript{D}\mu_{\mathrm{YMCS}}\right]\,\mathrm{e}^{-\Sigma}\,,
\label{ymcsmag7}
\end{equation}
with $\Sigma = S_{\mathrm{YMCS}}+S^{\mathrm{MAG}}_{\mathrm{FP}}$ and $\left[\EuScript{D}\mu_{\mathrm{YMCS}}\right]=\left[\EuScript{D}A\right]\left[\EuScript{D}b\right]\left[\EuScript{D}\bar{c}\right]\left[\EuScript{D}c\right]$. The tree-level propagators for the non-Abelian and diagonal gauge fields are, respectively,
\begin{equation}
\langle A^{a}_{\mu}(k)A^{b}_{\nu}(-k)\rangle = \frac{\delta^{ab}}{k^2+M^2}\left(\delta_{\mu\nu}-\frac{k_{\mu}k_{\nu}}{k^2}+M\frac{k_\rho}{k^2}\epsilon_{\mu\rho\nu}\right)\,,
\label{ymcsmag8}
\end{equation}
and
\begin{equation}
\langle A_{\mu}(k)A_{\nu}(-k)\rangle = \frac{1}{k^2+M^2}\left(\delta_{\mu\nu}-\frac{k_{\mu}k_{\nu}}{k^2}+M\frac{k_\rho}{k^2}\epsilon_{\mu\rho\nu}\right)\,.
\label{ymcsmag9}
\end{equation}
Expressions \eqref{ymcsmag8} and \eqref{ymcsmag9} show that both non-Abelian and Abelian components of the gauge field acquire the same non-vanishing topological mass at tree-level. The propagators are transverse, but such a property should not hold exactly at higher order to the non-Abelian sector due to the non-linear gauge condition \eqref{ymcsmag6}.

It is a well-known fact that the MAG is plagued by the Gribov problem see, e.g., \cite{Capri:2005tj}. Thanks to the fact that the Faddeev-Popov operator $\mathcal{M}^{ab}$ is Hermitian in this gauge, the definition of a Gribov region for the MAG is possible, see \cite{Capri:2005tj,Dudal:2006ib,Capri:2006cz,Capri:2008ak,Capri:2008vk,Capri:2010an,Guimaraes:2011sf,Capri:2013vka,Gongyo:2013rua}, and thereby the implementation of the restriction of the functional integral to such a region renders a partial solution to the Gribov problem. In fact, the Gribov region is free of infinitesimal Gribov copies and, in the present case, due to the Chern-Simons term, gauge invariance is verified just for infinitesimal transformations. Thence, for Yang-Mills-Chern-Simons theory, removing infinitesimal Gribov copies actually corresponds to deal with the Gribov problem completely. Invariance under finite gauge transformation is achieved only if the mass parameter $M$ is chosen to satisfy a quantization rule \cite{Deser:1981wh,Deser:1982vy}. 
\section{Getting rid of Gribov copies}

In four-dimensional Yang-Mills theories, infinitesimal Gribov copies were eliminated in the MAG by the restriction of the functional integral to the correspondent Gribov region in the MAG, see \cite{Capri:2008vk}. Such a region is defined as
\begin{equation}
\Omega_{\mathrm{MAG}} = \left\{(A^{a}_{\mu},A_\mu),\EuScript{D}^{ab}_{\mu}A^{b}_{\mu}=0\,,\partial_\mu A_\mu=0\,|\,\mathcal{M}^{ab}>0\right\}\,.
\label{grgc1}
\end{equation}  
It features important properties to be a good candidate for the restriction of the functional integral. In particular, the region is bounded in field space in the non-Abelian directions while it is unbounded in the diagonal ones. The referred boundary is again dubbed Gribov horizon. Furthermore, for every configuration close to the horizon, there is a copy localized outside the horizon, suggesting that the Gribov region does not exclude any physical configuration \cite{Capri:2008vk}. As in the Landau gauge, the restriction of the path integral to $\Omega_{\mathrm{MAG}}$ can be achieved by the so-called Gribov no-pole condition. This ammounts to impose that the only pole developed by the ghost propagator is the trivial $p^2=0$ one. In the case of the MAG, the no-pole condition is imposed to the non-Abelian ghosts. This was worked out before in pure Yang-Mills theories, see \cite{Capri:2005tj}. However, the restriction does not refer to the dynamical action in the partition function and therefore, the same procedure can be directly imported to the Yang-Mills-Chern-Simons path integral with the only difference that such a theory is defined in three dimensions. The restriction is a geometric procedure and can be easily worked out for general spacetime dimension.

Formally, the restriction is given by the modification of the measure as
\begin{equation}
\EuScript{Z}_{\mathrm{MAG}}=\int\left[\EuScript{D}\mu_{\mathrm{YMCS}}\right]\EuScript{V}(\Omega_{\mathrm{MAG}})\,\mathrm{e}^{-\Sigma}\,,
\label{grgc2}
\end{equation}
where $\EuScript{V}(\Omega_{\mathrm{MAG}})$ works as a cutoff at the Gribov horizon, i.e., the region where the Faddeev-Popov operator hits the first zero modes. According to the Gribov no-pole prescription, one has to evaluate the connected ghost two-point function by considering the gluon as an external field. In the case of the MAG, we concentrate on the non-Abelian ghost propagator. The restriction is imposed by demanding that the resulting two-point function does not develop poles besides $p^2=0$. Therefore, at leading order in the coupling $g$, the two-point function is
\begin{equation}
\EuScript{G}(p) = \frac{1}{2V}\sum_{a=1,2}\langle\bar{c}^a (p) c^a(-p)\rangle_{A}\,,
\label{grgc3}
\end{equation}
where $V$ stands for the spacetime volume and $\langle\ldots \rangle_A$ denotes that the correlation function is computed by taking $A$ as an external field. Explicitly, at leading order,
\begin{equation}
\EuScript{G}(p) = \frac{1}{p^2}(1+\sigma (p,A)) + \frac{B}{p^4}\,,
\label{grgc4}
\end{equation}
where 
\begin{equation}
\sigma (p,A) = \frac{4g^2}{3V}\int \frac{\mathrm{d}^3k}{(2\pi)^3}\frac{A_{\alpha}(k)A_{\alpha}(-k)}{(p-k)^2}\,,
\label{grgc5}
\end{equation}
and $B$ is independent of $p$ and it is positive.  Therefore, just $\sigma(p,A)$ can generate a non-trivial pole than $p^2=0$ and the terms containing $B$ are actually not important for our discussion. Thus, we rewrite \eqref{grgc4} as 
\begin{equation}
\EuScript{G}(p) \approx \frac{1}{p^2}\frac{1}{1-\sigma (p,A)}\,,
\label{grgc6}
\end{equation}
and the non-trivial pole is avoided as long as $\sigma (p,A)<1$. It can be shown that $\sigma (p,A)$ monotonically decreases with $p^2$. Hence, a $p$-independent way of imposing that no non-trivial poles are generated corresponds to demand,
\begin{equation}
\sigma(0,A)<1\,,
\label{grgc7}
\end{equation}
which is referred to as the no-pole condition. Thence, the function $\EuScript{V}(\Omega_{\mathrm{MAG}})$ which modifies the path integral measure can be chosen to be a Heaviside step function of the form
\begin{equation}
\EuScript{V}(\Omega_{\mathrm{MAG}})=\theta (1-\sigma(0,A))\,.
\label{grgc8}
\end{equation}
At this point, the reader can appreciate that the form of the modification to the functional measure is completely independent of the fact the we are dealing with Yang-Mills-Chern-Simons theory rather than pure Yang-Mills simply because the same operator, i.e., the Faddeev-Popov operator, defines the Gribov problem in both cases and therefore the same set of zero-modes must be removed. Using the integral representation of the $\theta$-function, i.e.,
\begin{equation}
\theta (1-\sigma(0,A)) = \int^{i\infty + \epsilon}_{-i\infty + \epsilon}\frac{\mathrm{d}\zeta}{2\pi i\zeta}\mathrm{e}^{\zeta (1-\sigma(0,A))}\,,
\label{grgc9}
\end{equation}
it is possible to lift the modification on the path integral measure to a Boltzmann factor, leading to an effective term in the action. Such a modification involves the factor $\sigma (0,A)$ which contains the Abelian gauge fields. As a consequence, it entails a modification to the gluon propagator. Retaining up to quadratic terms in the partition function and integrating out all fields leads to
\begin{eqnarray}
\EuScript{Z}^{\mathrm{quad}}_{\mathrm{MAG}} &=&\lim_{\alpha,\beta\to 0}\EuScript{N}\int\frac{\mathrm{d}\zeta}{2\pi i}\mathrm{e}^{\zeta-\mathrm{ln}~\zeta}\,\mathrm{det}^{-1/2}\Delta^{ab}_{\mu\nu}\nonumber\\
&\times&
\mathrm{det}^{-1/2}\tilde{\Delta}_{\alpha\beta}\,,
\label{grgc10}
\end{eqnarray}
 with $\EuScript{N}$ being a normalization factor. The operators $\Delta$ and $\tilde{\Delta}$ are
 \begin{equation}
 \Delta^{ab}_{\mu\nu} = \delta^{ab}\left[\delta_{\mu\nu}p^2-\left(1+\frac{1}{\alpha}\right)p_\mu p_\nu -M\epsilon_{\mu\rho\nu}p_\rho\right]\,,
 \label{grgc11}
 \end{equation}
 and
 \begin{equation}
 \tilde{\Delta}_{\mu\nu} = \delta_{\mu\nu}\left(p^2+\frac{8g^2}{3V}\frac{\zeta}{p^2}\right)-\left(1+\frac{1}{\beta}\right)p_\mu p_\nu -M\epsilon_{\mu\rho\nu}p_\rho\,.
 \label{grgc12}
 \end{equation}
The determinant of the operator $\Delta^{ab}_{\mu\nu}$ can be absorbed in the normalization factor by defining $\EuScript{N}^{\prime}\equiv\EuScript{N}\,\mathrm{det}^{-1/2}\Delta^{ab}_{\mu\nu}$. Following the standard strategy, the remaining integral is evaluated in a saddle-point approximation,
\begin{equation}
\EuScript{Z}^{\mathrm{quad}}_{\mathrm{MAG}}  = \EuScript{N}^{\prime}\,\mathrm{e}^{f(\zeta^\ast)}\,,
\label{grgc13}
\end{equation}
with
\begin{equation}
f(\zeta)=\zeta-\mathrm{ln}~\zeta-\frac{1}{2}\mathrm{Tr\,ln}~\tilde{\Delta}_{\mu\nu}\,,
\label{grgc14}
\end{equation}
and $\zeta^\ast$ is the solution of
\begin{equation}
\frac{\partial f(\zeta)}{\partial \zeta}\Big|_{\zeta=\zeta^\ast}=0\,.
\label{grgc15}
\end{equation}
Therefore,
\begin{equation}
1-\frac{1}{\zeta^\ast}-\frac{1}{2}\mathrm{Tr}\left[\frac{\partial \tilde{\Delta}_{\mu\alpha}}{\partial \zeta}\tilde{\Delta}^{-1}_{\alpha\nu}\right]\Bigg|_{\zeta=\zeta^\ast}=0\,.
\label{grgc16}
\end{equation}
Upon explicit evaluation of the last term and taking the limits $V\to\infty$ and $\beta\to 0$ while holding $\gamma^4 \equiv \frac{8g^2\zeta^\ast}{3V}$ finite yields
\begin{equation}
\frac{8g^2}{3}\int \frac{\mathrm{d}^3 p}{(2\pi)^3}\frac{p^4+\gamma^4}{M^2p^6+(p^4+\gamma^4)^2}=1\,.
\label{grgc17}
\end{equation}
This gap equation fixes $\gamma$, from now on referred to as the Gribov parameter. Such a parameter $\gamma$ is dimensionful and will set a mass scale to the propagators of the theory. The gap equation also contains the topological mass $M$, and reduces to the standard expression in pure Yang-Mills theories when $M\to 0$.  In three dimensions, the gap equation can be directly solved by performing the integral in \eqref{grgc17} which is convergent. The explicit expression is not particularly illuminating and therefore, we do not write it here.

With the Gribov parameter fixed by the gap equation \eqref{grgc17}, we can recompute the tree-level non-Abelian and Abelian gluon propagators. From eq.\eqref{grgc11} and \eqref{grgc12} it is clear that the Abelian propagator is affected by the presence of the Gribov horizon. Explicitly, the Abelian propagator is
\begin{eqnarray}
\langle A_{\mu}(p)A_{\nu}(-p)\rangle &=&\frac{p^2 (p^4 +\gamma^4)}{(p^4+\gamma^4)^2+M^2 p^6}\left(\delta_{\mu\nu}-\frac{p_\mu p_\nu}{p^2}\right.\nonumber\\
&+&\left.\frac{p^2}{p^4+\gamma^4}M\epsilon_{\mu\lambda\nu}p_\lambda\right)
\label{grgc18}
\end{eqnarray}
where there is a clear modification with respect to the tree-level expression when the restriction to the Gribov horizon is not taken into account, see eq.\eqref{ymcsmag9}. The non-Abelian propagator, at tree-level, is the same as in eq.\eqref{ymcsmag8}. In the limit $M\to 0$, one recovers the same propagators of Yang-Mills theories quantized in the MAG and restricted to the Gribov horizon, see \cite{Capri:2005tj}. The effects of the Gribov parameter to the pole structure of the gluon propagator will be discussed in the next Section.

The restriction of the path integral to the Gribov region through the no-pole condition was implemented at leading order here. This follows the original strategy developed by Gribov \cite{Gribov:1977wm} in the Landau gauge. In \cite{Capri:2012wx}, it was shown that, in the Landau gauge, the no-pole condition, when implemented at all orders in perturbation theory, is equivalent to Zwanziger's horizon condition \cite{Zwanziger:1989mf}. Although such equivalence was not formally established in the MAG so far, the restriction to the Gribov horizon in the MAG was implemented using Zwanziger's method in \cite{Gongyo:2013rua}. The result agrees with previous studies as in \cite{Capri:2005tj,Capri:2006cz}, where the choice of the precise form of the horizon function was constrained by renormalizability and localizability. 

Taping on that, we are now able to write the Gribov-Zwanziger action associated to Yang-Mills-Chern-Simons theories quantized in the MAG. It is expressed as,
\begin{equation}
S^{\mathrm{nloc}}_{\mathrm{GZ}} = \Sigma + S_{\mathrm{H}} + \mathrm{Vol}\,,
\label{grgc19}
\end{equation}
with\footnote{We have performed a redefinition of the Gribov parameter $\gamma^2\to g\gamma^2$ and $\mathrm{Vol}$ is a volume term whose explicit form is irrelevant for the present purposes.}
\begin{equation}
S_{\mathrm{H}} = g^2\gamma^{4}\int_{x,y}\epsilon^{ab}A_{\mu}(x)\left[\mathcal{M}^{-1}\right]^{ac}(x,y)\,\epsilon^{cb}A_{\mu}(y)\,,
\label{grgc20}
\end{equation}
being the horizon function. As discussed in Sect.~\ref{pre}, the non-locality introduced by the horizon function can be localized by the introduction of auxiliary fields. The local Gribov-Zwanziger action $S^{\mathrm{MAG}}_{\mathrm{GZ}}$ associated to Yang-Mills-Chern-Simons theories in the MAG is written as
\begin{eqnarray}
S^{\mathrm{MAG}}_{\mathrm{GZ}} &=& \Sigma - \int_x \left(\bar{\varphi}^{ac}_{\mu}\mathcal{M}^{ab}\varphi^{bc}_{\mu}-\bar{\omega}^{ab}_{\mu}\mathcal{M}^{ab}\omega^{bc}_{\mu}\right)\nonumber\\
&+& \gamma^2\int_x g\epsilon^{ab}A_\mu (\bar{\varphi}+\varphi)^{ab}_{\mu} + \mathrm{Vol}\,.
\label{grgc21}
\end{eqnarray}
In the case where the topological mass is absent $(M=0)$, action \eqref{grgc21} was shown to be renormalizable at all orders in perturbation theory in four dimensions, \cite{Capri:2006cz}. The renormalizability properties of \eqref{grgc21} in the presence of the Chern-Simons term will be reported elsewhere. 

At this stage, one can discuss many different formal aspects regarding action \eqref{grgc21}. In particular, the introduction of the horizon function as described yields a soft breaking of BRST symmetry. The breaking is soft since in the deep ultraviolet, the Gribov parameter vanishes and BRST invariance is restored. Such a breaking was vastly studied in the literature \cite{Dudal:2009xh,Sorella:2009vt,Baulieu:2008fy,Capri:2010hb,Dudal:2012sb,Dudal:2014rxa,Pereira:2013aza,Pereira:2014apa,Capri:2014bsa,Serreau:2012cg,Lavrov:2013boa,Moshin:2014xka,Schaden:2014bea,Schaden:2015uua,Cucchieri:2014via}. Nevertheless, as pointed out in \cite{Capri:2015ixa} in linear covariant gauges and generalized to the MAG in \cite{Capri:2015pfa}, a manifest BRST-invariant formulation of the horizon function is possible. In this case, gauge-invariant ``dressed" fields replace the gauge fields in the horizon function. However, instead of elaborating on that, we will focus on the analytic structure of the gluon propagator in the next section.

\section{Non-perturbative Abelian gluon propagator: analytic structure}

The restriction of the path integral measure to the Gribov region engendered an effective, non-perturbative modification of the Abelian gluon propagator. As such, its analytic structure is affected and thereby a discussion about the spectrum of the theory is deserved. In particular, from \eqref{grgc18}, one sees that the propagator develops poles due to the vanishing of the function $\EuScript{F}(p)$, defined as
\begin{equation}
\EuScript{F}(p) \equiv (p^4 +\gamma^4)^2 + M^2 p^6 = p^8 + \gamma^8 + 2p^4 \gamma^4 + M^2 p^6\,. 
\label{npgpas1}
\end{equation}
It is convenient to parameterize eq.\eqref{npgpas1} as 
\begin{equation}
\EuScript{F}(p) = (p^2+m^2_1)(p^2+m^2_2)(p^2+m^2_3)(p^2+m^2_4)\,,
\label{npgpas2}
\end{equation}
with $(m^2_1,m^2_2,m^2_3,m^2_4)$ standing for the roots of \eqref{npgpas1}. Decomposing the propagator \eqref{grgc18} into parity-preserving and violating pieces as
\begin{equation}
\langle A_{\mu}(p)A_{\nu}(-p)\rangle = \EuScript{K}^{\mathrm{pres}}_{\mu\nu}(p) + \EuScript{K}^{\mathrm{viol}}_{\mu\nu}(p)\,,
\label{npgpas3}
\end{equation}
with
\begin{equation}
\EuScript{K}^{\mathrm{pres}}_{\mu\nu}(p) = \frac{p^2 (p^4 +\gamma^4)}{(p^4+\gamma^4)^2+M^2 p^6}\left(\delta_{\mu\nu}-\frac{p_\mu p_\nu}{p^2}\right)\,,
\label{npgpas4}
\end{equation}
and
\begin{equation}
\EuScript{K}^{\mathrm{viol}}_{\mu\nu}(p) = \frac{M p^4}{(p^4+\gamma^4)^2+M^2 p^6}\epsilon_{\mu\lambda\nu}p_\lambda\,.
\label{npgpas5}
\end{equation}
Each sector of the propagator is written in a partial-fraction like decomposition, leading to
\begin{equation}
\EuScript{K}^{\mathrm{pres}}_{\mu\nu}(p) = \sum_{i=1}^{4}\frac{E_i}{p^2+m^2_i}\left(\delta_{\mu\nu}-\frac{p_\mu p_\nu}{p^2}\right)\,,
\label{npgpas6}
\end{equation} 
where
\begin{equation}
E_i =\frac{m^2_i (m^4_i + \gamma^4)}{\prod_{j=1\,,\, j\neq i}^{4}(m^2_i-m^2_j)}\,.
\label{npgpas7}
\end{equation}
For the parity-violating sector, the decomposition is expressed as
\begin{equation}
\EuScript{K}^{\mathrm{viol}}_{\mu\nu}(p) = \sum_{i=1}^{4}\frac{B_i}{p^2+m^2_i}\epsilon_{\mu\lambda\nu}p_\lambda\,,
\label{npgpas8}
\end{equation}
with
\begin{equation}
B_i = -\frac{M^4 m^4_i}{\prod_{j=1\,,\, j\neq i}^4 (m^2_i - m^2_j)}\,.
\label{npgpas9}
\end{equation}
Written as in \eqref{npgpas6} and \eqref{npgpas8}, the pole and residue structure is made manifest. Those quantities depend on the coupling constant $g$, on the Gribov parameter $\gamma$ and on the Chern-Simons mass $M$. Such parameters are correlated by the gap equation \eqref{grgc17}. Therefore, a complete analysis of the analytic structure of the corresponding propagator requires the solution of the gap equation at some given order in perturbation theory. In this work, however, similarly to \cite{Canfora:2013zza}, we treat those parameters as being free and characterize the spectrum of the theory for arbitrary values of them. Although this strategy gives too much freedom for the allowed values of each parameter, it does not rely on a specific solution of the gap equation at a given order in pertubation theory, a fact which might allow to mimic more refined results in a perturbative expansion. 

\begin{figure}[htp]
	\begin{center}
		\includegraphics[width=8.5cm]{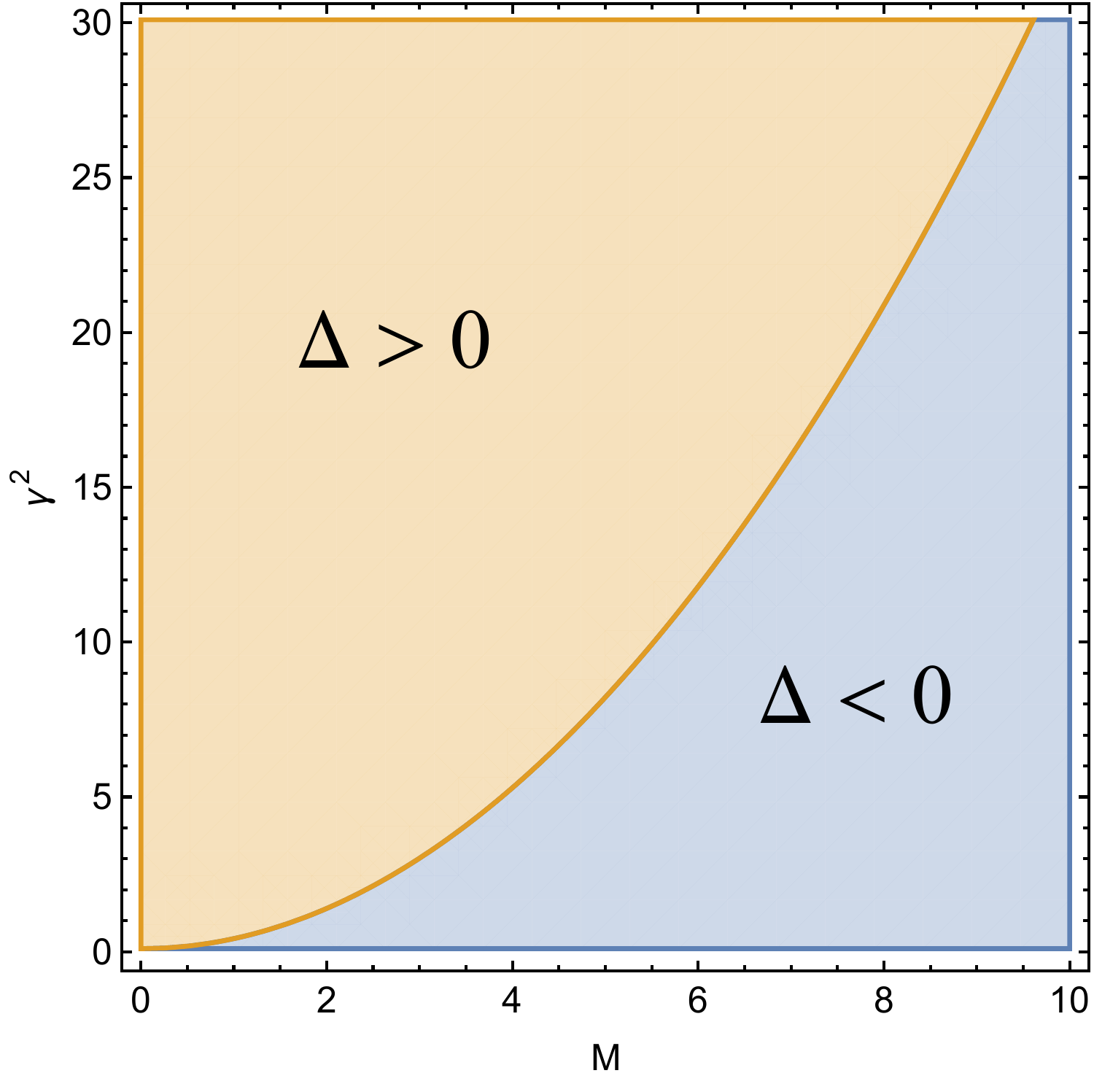} 
		\caption{\footnotesize Sign of the discriminant $\Delta$ of $\EuScript{F}(p)$ as a function of $M$ and $\gamma^2$.}
		\label{discriminant}
	\end{center}
\end{figure}

In order to characterize the nature of the excitations, we have to determine the pole structure of the propagator. In particular, we have to find the roots of $\EuScript{F}(p)$. Redefining $p^2\to\bar{p}$ we obtain the quartic equation,
\begin{equation}
\bar{p}^4+M^2\bar{p}^3+2\gamma^4\bar{p}^2+\gamma^8 = 0\,.
\label{npgpas10}
\end{equation}
By employing the standard definition of the discriminant $\Delta$ for a quartic equation, one gets
\begin{equation}
\Delta = 256M^4 \gamma^{20} - 27M^8 \gamma^{16}\,.
\label{npgpas11}
\end{equation}
The pole structure depends on the sign of the discriminant $\Delta$. In Fig.~\ref{discriminant} we plot the sign of $\Delta$ as a function of $\gamma^2$ and $M$. 
We now analyze each situation separately. If $\Delta > 0$, it is a known fact that either the poles are all real or all complex. This is established by the sign of subsidiary polynomials defined by
\begin{equation}
\EuScript{P}(M,\gamma) = 16\gamma^4-3M^4\,,
\label{npgpas111}
\end{equation}
and
\begin{equation}
\EuScript{D}(M,\gamma) = 32M^4\gamma^4 - 3M^8\,.
\label{npgpas12}
\end{equation}
If $\EuScript{P}(M,\gamma)<0$ and $\EuScript{D}(M,\gamma)<0$, then all roots are real. It turns out that for the values of $\gamma^2$ and $M$ that we use in the parameter space, there is no overlap of regions where all such conditions are simultaneously satisfied. Hence, no real roots are found. However if $\EuScript{P}(M,\gamma)>0$ or $\EuScript{D}(M,\gamma)>0$ with $\Delta > 0$, then all roots are complex. The overlapping region where those conditions are satisfied coincides with the region where $\Delta>0$ in Fig.~\ref{discriminant}. Therefore, for a wide range of values of $(\gamma^2,M)$ all poles are complex and therefore, all excitations cannot be part of the physical spectrum. This can be interpreted as a signal of confinement. 

\begin{figure}[htp]
	\begin{center}
		\includegraphics[width=8.5cm]{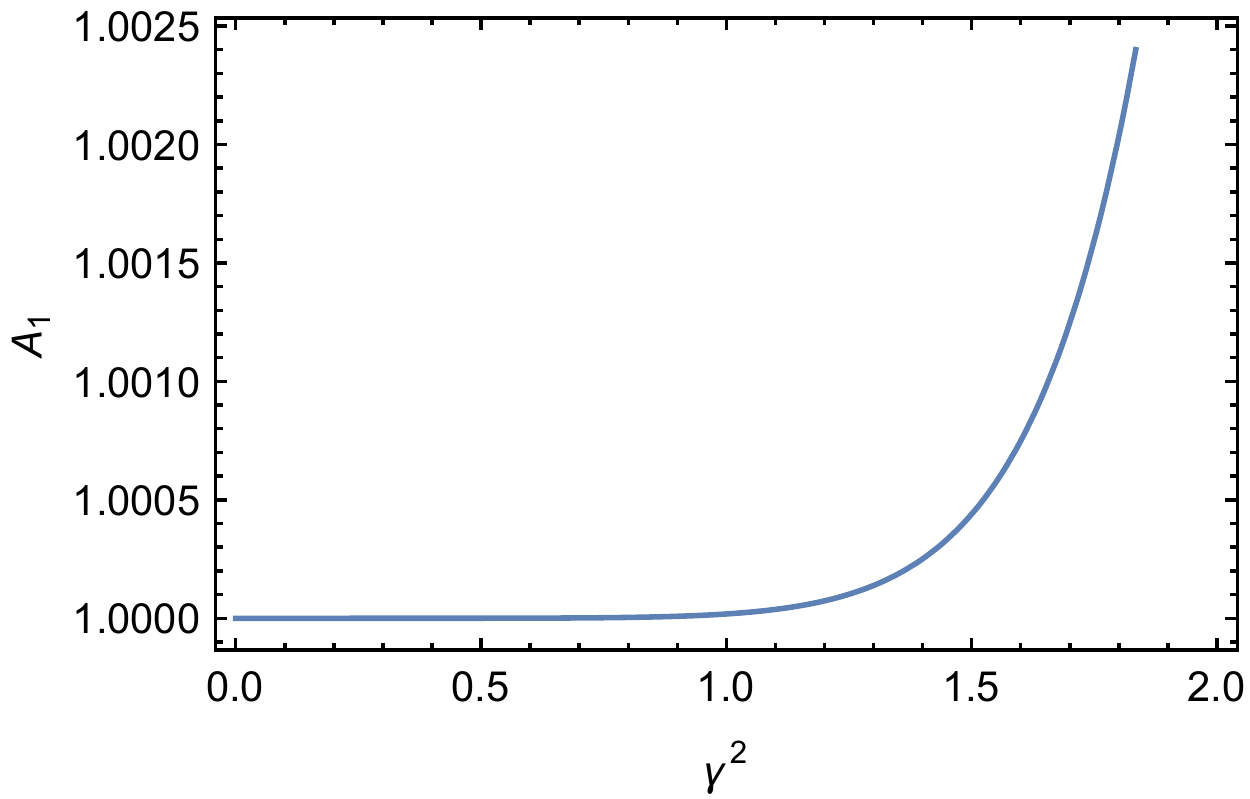} 
		\caption{\footnotesize Residue $A_1$ - associated to the parity-preserving part of the propagator - for $M=-5$.}
		\label{residueA1}
	\end{center}
\end{figure}

\begin{figure}[htp]
	\begin{center}
		\includegraphics[width=8.5cm]{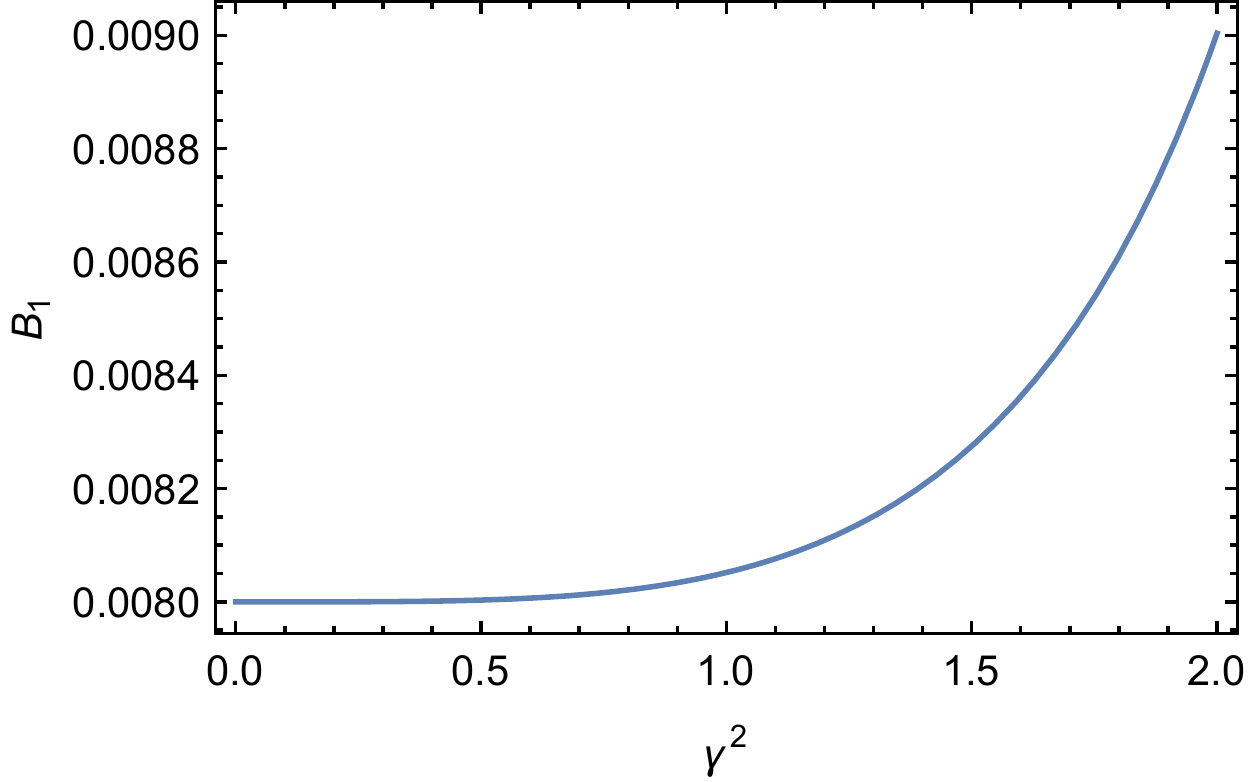} 
		\caption{\footnotesize Residue $B_1$ - associated to the parity-violating part of the propagator - for $M=-5$.}
		\label{residueB1}
	\end{center}
\end{figure}

\begin{figure}[htp]
	\begin{center}
		\includegraphics[width=8.5cm]{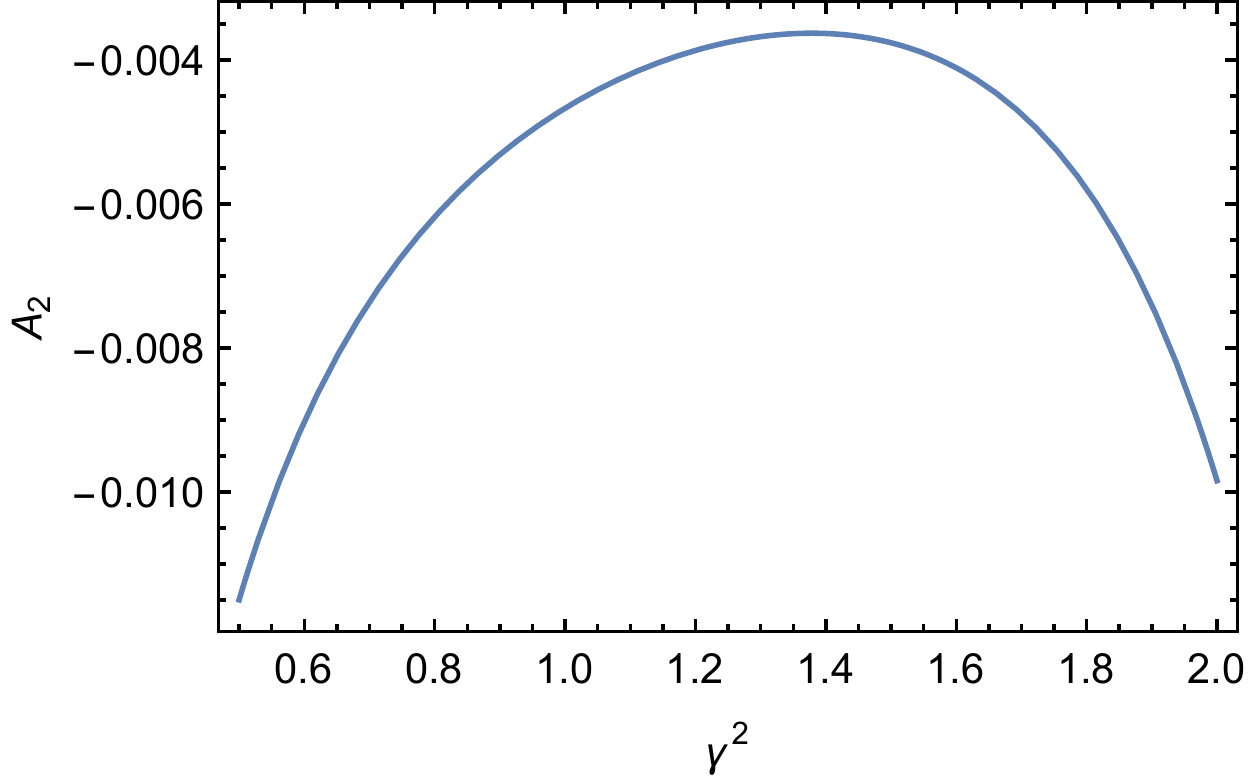} 
		\caption{\footnotesize Residue $A_2$ - associated to the parity-preserving part of the propagator - for $M=-5$.}
		\label{residueA2}
	\end{center}
\end{figure}

\begin{figure}[t!]
	\begin{center}
		\includegraphics[width=8.5cm]{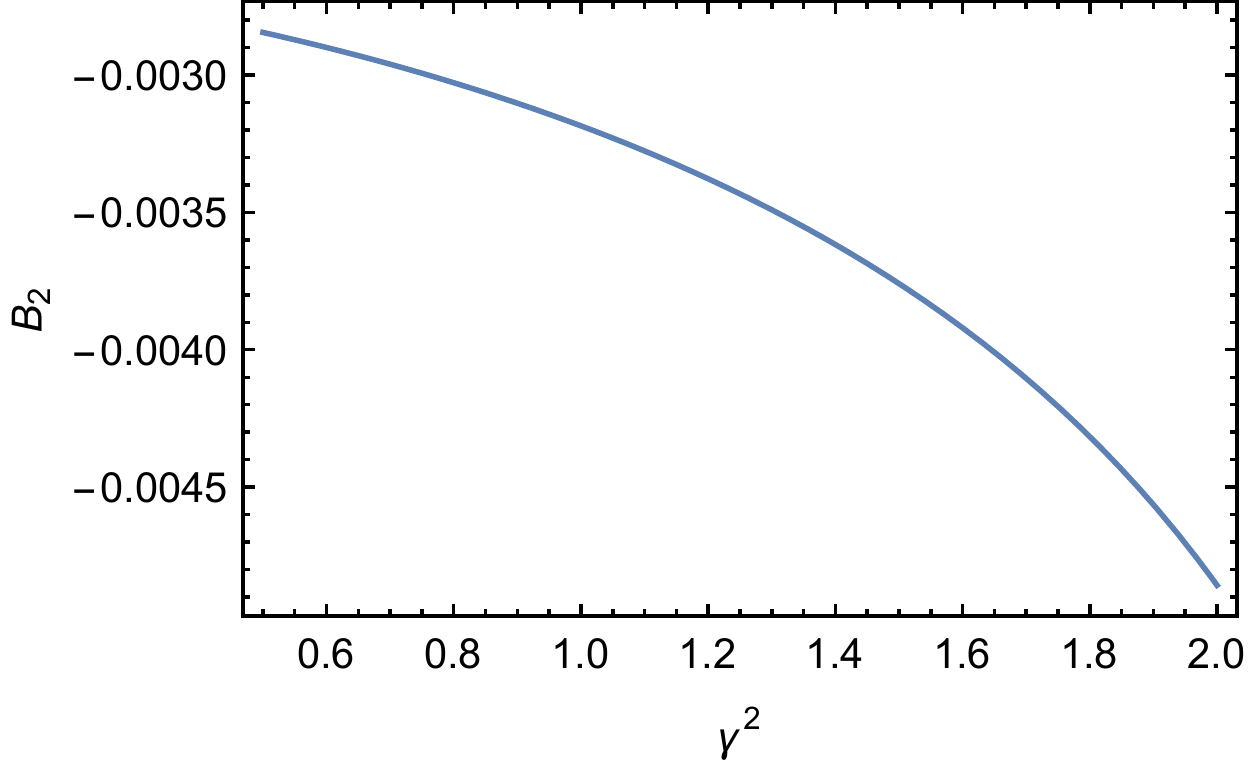} 
		\caption{\footnotesize Residue $B_2$ - associated to the parity-violating part of the propagator - for $M=-5$.}
		\label{residueB2}
	\end{center}
\end{figure}

On the other hand, if $\Delta < 0$ then $\EuScript{F}$ has two distinct real roots and two complex conjugate roots. In Fig.~\ref{discriminant}, the region where $\Delta < 0$ is indicated. Having two real roots, it is conceivable that physical excitations can be generated. This is determined by the sign of the residues. The poles of the parity-preserving and violating parts are the same. For $\Delta < 0$, it is possible to find suitable values of $\gamma^2$ and $M$ for which the residues associated to one of the real poles is positive and therefore can be associated to a physical excitation. Hence, by changing the values of those ``free" parameters the theory exhibits different regimes: a confining one, in the sense that all poles are complex and cannot be associated to physical excitations in the spectrum and a different one where a physical excitation can show up. For concreteness, we plot in Fig.~\ref{residueA1} and \ref{residueB1} the values of the residues for the parity-preserving and violating parts of the propagator.
For the other real pole, the residues are negative and therefore, it cannot be associated to a physical excitation. The residues for $M=-5$, as a function of $\gamma^2$ are plotted in Fig.~\ref{residueA2} and \ref{residueB2}.
Hence, there is a a region in parameter space where the discriminant $\Delta$ is negative and two real poles are generated. One of them has positive residues and can be associated to a physical excitation while the other real pole does not feature positive residues as shown in Fig.~\ref{residueA2} and \ref{residueB2}. Then, qualitatively, the theory displays two different phases in the parameter space: One in which the discriminant $\Delta$ is positive and all poles are complex being thus interpreted as confined excitations. The other region where $\Delta$ is negative admits two real poles, one of which with positive residues, being thus interpreted as a deconfined excitation. Such possibilities arise due to the interplay of the mass parameters $M$, which has a topological nature and $\gamma^2$, which arises from the restriction of the path integral to the Gribov region. A similar behavior was observed in Yang-Mills-Higgs systems, see \cite{Capri:2012cr,Capri:2012ah,Capri:2013gha,Capri:2013oja} and Yang-Mills-Chern-Simons-Higgs \cite{Gomez:2015aba} in the Landau gauge. Inhere, such a transition from confined to deconfined phases is verified just in the Abelian sector which, as previously discussed, is the one affected by the restriction to the Gribov region. In the case of the non-Abelian sector, the absence of competing mass parameters does not allow for such a transition from confined to deconfined phases. Hence, the non-Abelian propagator has a Yukawa-like behavior. According to the Abelian-dominance hypothesis, such a mass  should be large enough in order to decouple those degrees of freedom in the infrared. 

\section{Perspectives and Conclusions}
In this work, we quantized the Yang-Mills-Chern-Simons theory in the MAG and took into account the existence of infinitesimal Gribov copies. Such a system features an intrinsic topological mass arising from the Chern-Simons term and a dynamical mass parameter - the Gribov parameter - associated to the restriction of the path integral measure to the Gribov region. As it is known, the restriction to the Gribov region affects the Abelian gluon propagator and thereby due to a suitable choice of the mass parameters, the Abelian gluon propagator displays real poles or purely complex poles. This is interpreted as a transition from a confining to a deconfining phase. It is important to emphasize that the Gribov parameter is not free, but fixed by a gap equation. In particular, it is determined in terms of the coupling constant $g$ and the topological mass $M$ as shown by the leading order contribution to the gap equation in eq.\eqref{grgc17}. In this contribution, we have treated the mass parameters as being free and concentrated the analysis to the viability of the the phase transition instead of trying to make it quantitative.

The Yang-Mills-Chern-Simons model is invariant just under infinitesimal transformations for generic values of the topolofical mass. Hence, in this particular case, eliminating only infinitesimal Gribov copies is in fact equivalent to remove all the Gribov copies - this is in contrast to standard Yang-Mills theories where within the Gribov region there are still large copies \cite{vanBaal:1991zw}. 

As it is known in the case of pure Yang-Mills theories, the Gribov-Zwanziger theory suffers from infrared instabilities which favors the formation of dimension-two condensates. This leads to the so-called Refined Gribov-Zwanziger action which takes into account those condensates from the begining \cite{Capri:2015pfa,Capri:2015nzw,Dudal:2008sp,Guimaraes:2015bra}. Such a feature was investigated in different gauges, including the MAG. In principle, in the Yang-Mills-Chern-Simons model, the generation of dimension-two condensates would also occur. In this case, the mass parameters associated to the condensates would have their own gap equations fixing them in terms of $g$ and $M$. Ultimately, this would affect the pole structure of the gluon propagator leading to a new phase diagram. An explicit check of the viability of a transition from confined to deconfined phases would be required. This is beyond the scope of this paper. 

The introduction of the horizon function as discussed in Sect.~IV breaks in an explicit but soft way the BRST symmetry, see \cite{Capri:2008ak}. However, as discussed in \cite{Capri:2015pfa} and \cite{Capri:2017abz} in the MAG, it is possible to provide a manifestly BRST-invariant formulation of the Gribov-Zwanziger action by introducing a gauge-invariant field $A^h_\mu$. This has an important consequence of providing a physical meaning for the Gribov parameter, i.e., it is not akin to a gauge parameter. This is a formal development that will be reported elsewhere. Finally, the renormalizability properties of the Yang-Mills-Chern-Simons theory quantized in the MAG restricted to the Gribov region will appear in a forthcoming publication.

\section*{Acknowledgments}
The authors are grateful to S. P. Sorella for discussions. ADP acknowledges ACRI under the Young Investigator Program and SISSA for hospitality, CNPq under the grant PQ-2 (309781/2019-1) and FAPERJ under the ``Jovem Cientista do Nosso Estado" program (E-26/202.800/2019). The Coordena\c c\~ao de Aperfei\c coamento de Pessoal de N\'ivel Superior (CAPES) is also acknowledged for financial support.

\appendix

\section{BRST transformations}\label{appendix}
For the sake of completeness, we list here the BRST transformations generated by the nilpotent operator $s$ $(s^2 = 0)$ for the non-Abelian and Abelian components of the field content of Yang-Mills-Chern-Simons theories quantized in the MAG with gauge group being $SU(2)$. They are,
\begin{eqnarray}
sA^a_\mu &=& -\EuScript{D}^{ab}_{\mu}c^b - g\epsilon^{ab}A^b_\mu c\,,\nonumber\\
sc^a &=& g\epsilon^{ab}c^b c\,,\nonumber\\
s\bar{c}^a &=& i b^a\,,\nonumber\\
sb^a &=& 0\,,\nonumber\\
sA_\mu &=& -\partial_\mu c - g\epsilon^{ab}A^a_\mu c^b\,,\nonumber\\
sc &=& = \frac{g}{2}\epsilon^{ab}c^a c^b\,,\nonumber\\
s\bar{c} &=& ib\,,\nonumber\\
sb&=& 0\,.
\label{ap1}
\end{eqnarray}
\bibliography{refsymcsmag}

\end{document}